\documentstyle[12pt,psfig,epsf]{article}
\newcommand{\be}{\begin{equation}}
\newcommand{\ee}{\end{equation}}
\newcommand{\pr}{\prime}
\newcommand{\pp}{\prime \prime}
\newcommand{\ppp}{\prime \prime \prime}
\newcommand{\pppp}{\prime \prime \prime \prime}
\newcommand{\p}{\partial}
\newcommand{\f}{\frac}

\begin {document}
\thispagestyle{empty}

\hfill NTUA-77/99

\begin {center}
{\Large {\bf The Abelian Higgs Model in Three Dimensions with
Improved Action}} 
\end{center}

\vspace{3cm}

\begin {center}

{\bf P.Dimopoulos, K.Farakos and G.Koutsoumbas}

\vspace{3cm}

Physics Department

National Technical University, Athens

Zografou Campus, 157 80 Athens, GREECE

\end{center}

\vspace{2cm}

\begin {center}
ABSTRACT
\end{center}

We study the Abelian Higgs Model using an improved form of the action in 
the scalar sector. The subleading corrections are carefully analysed and 
the connection between lattice and continuous parameters is worked out.
The simulation shows a remarkable improvement of the numerical
performance.

\vfill

\newpage

\section{Introduction}

The Abelian Higgs Model has mostly been studied in recent years as a
theoretical laboratory in the context of the Electroweak Baryogenesis 
scenario. As it is well known by now, the lattice investigations of the 
model are very demanding in computer power and time. It would be 
helpful to use an improved form of the lattice action, 
to reduce the autocorrelation and come closer to the continuum limit. 

We did not try an improvement on the pure gauge sector of the action. The 
strategy which is usually followed does not seem very reliable: actually
it leaves behind several subleading contributions, so it is not very
efficient in eliminating all of the unwanted terms. 
On the other hand, the weak coupling
regime where these simulations are performed suggest that the improvement 
of the gauge sector may not be very important. 
We mainly concentrated on improving the scalar sector; 
the procedure suffers more or less from the same
problems, however even a modest improvement is very important in this case.
The model on which we will work has been treated before (\cite{dimo}), so
we have reference results to compare with.

A promising approach to study the 
four-dimensional model at finite temperature is through reduction 
to an effective model in three dimensions.
This can be done if the couplings are small
and the temperature is much larger than any other mass scale in
the theory. The parameters of the reduced theory are
related to the ones of the original model through perturbation
theory. The reduced theory has some advantages over the
original one from the computational point of view. It is
super-renormalizable and yields transparent relations between
the (dimensionful) continuous parameters and the lattice ones. Moreover,
the number of mass scales is drastically reduced: (a) the scale $T,$
present in four dimensions is evidently absent, (b) one may also
integrate out the temporal component $A_0$ of the gauge field,
so its mass scale $g T$ also
disappears. Thus there are two mass scales less and this
reduces substantially the computer time needed to get reliable results.

The action for the U(1) gauge--Higgs model at finite temperature is:
\be
S[A_{\mu}(\tau ,\bar x) , \varphi (\tau ,\bar x)]=\int_{0}^{\beta } d\tau
\int d^3x [\frac{1}{4}F_{\mu\nu}F_{\mu\nu} +|D_{\mu }\varphi |^2+
m^2\varphi ^*\varphi+\lambda (\varphi ^* \varphi)^2], \ee
where $\beta = 1/T.$

If the action is expressed in terms of Fourier components, the mass terms are
 of the type:
\be
[(2\pi nT)^2 + (\vec k)^2]|A_\mu(n,\vec k)|^2
\ee
\be
[(2\pi nT)^2 + (\vec k)^2]|\varphi (n,\vec k)|^2
\ee
where $n=-\infty , \dots, \infty.$

At high temperatures T and energy scales less than $2\pi T$ the non--static modes 
$A_\mu(n\neq0,\vec k),~~\varphi(n\neq0,\vec k)$ are then supressed by the factor
$(2\pi nT)^2$ relative to the static $A_\mu(n=0,\vec k)$ and $\varphi(n=0,\vec k)$
modes. The method of dimensional reduction consists in integrating out the non--static
modes in the action and deriving an effective action (\cite {krs}). 
We notice that the mass of the adjoint Higgs field is of order $g T,$ which is large 
compared to $g^2 T,$ the typical scale of the theory. Thus, we integrate it out using 
perturbation theory (\cite {fkrs}). 

The effective action may then be written in the form: 
\be
S_{3D~eff}[A_{i}(\vec x),\varphi_{3}(\vec x)]=\int
d^3x[\frac{1}{4}F_{ij}F_{ij} +|D_{i}\varphi_3|^2+m_3^2\varphi_3^*
\varphi_3 +\lambda_3 (\varphi^*_3 \varphi_3)^2] \label{seff}
\ee

The index 3 in (\ref{seff}) is for the 3D character of the theory.
The relations between 4D and 3D parameters are (up to 1--loop):

\be g_{3}^{2}=g^{2}(\mu)T, \ee

\be
\lambda_{3}=T\lambda(\mu ) -\frac{g_{3}^{4}}{8\pi m_{D}},
\ee

$$m_{3}^{2}(\mu _{3})=\frac{1}{4}g_{3}^{2}T+\frac{1}{3}(\lambda _{3}+
\frac{g_{3}^{4}}{8\pi m_{D}})T-\frac{g_{3}^{2}m_{D}}{4\pi}-\frac{1}{2}m_{H}^{2}$$

\be
m_{D}^{2}=\frac{1}{3}g^{2}(\mu )T^{2}.
\ee

It is convenient to use the new set of parameters $(g_{3}^{2}, x,y)$ rather than the 
set $(g_{3}^{2}, \lambda_{3}, \mu_{3}^{2})$. $x,y$ are defined as (\cite {klrs}):
\be
x=\frac{\lambda_{3}}{g_{3}^{2}} \ee
\be y=\frac{m_{3}^{2}(g_{3}^{2})}{g_{3}^{4}} \ee

It is evident that $x$ is just proportional to the ratio of the squares of the scalar
and vector masses; $y$ is related to the temperature. The parameters $x,y$ can be expressed 
in terms of the four--dimensional parameters as follows(\cite {karjal}):

\be 
x=\frac{1}{2}\frac{m_{H}^{2}}{m_{W}^{2}}-\frac{\sqrt{3}\ g}{8\pi }
\label{mass}
\ee

\be
y=\frac{1}{4g^{2}}+\frac{1}{3g^{2}}(x+\frac{\sqrt{3}\ g}{8\pi })-\frac{1}{4\pi \sqrt{3}g}   
-\frac{m_{H}^{2}}{2 g^4 T^2}
\label{yT}
\ee

We have  concentrated on the phase transition line. We
have chosen to fix the Higgs mass to a fixed value (30 GeV),
g to 1/3, $m_W$ to 80.6 GeV and study the characteristics of
the phase transition.

\section{The improved lattice action}

The whole idea of improving the lattice action has been put
forward (\cite{symanzik}, \cite{lepage}, \cite{paf}, \cite{piroth}, 
\cite{weisz}) to enhance performance of the lattice calculations.
The lattice actions, when expanded in powers of the 
lattice spacing $a,$ yield the terms of the continuum action plus 
subleading terms, i.e. terms multiplied by higher powers of $a.$
The procedure is to include additional terms in the action, so that the
corrections that remain in the na\"ive continuum limit start at a 
higher power of the lattice spacing, as compared to the usual action.
This work follows most closely \cite{weisz}.

\subsection{The pure gauge part}

The pure gauge part is expressed by the plaquette term in the action. 
We use the non-compact formulation; the initial action is defined by
$\beta _{g}\sum _{x}\sum _{0<i<j} F_{ij}^{2}$, where $F_{ij} \equiv 
\Delta _{i}^f A_{j}(x)-\Delta _{j}^f A_{i}(x),~~\Delta _{i}^f A_{j}(x) \equiv
 A_j(x+{\hat i}) - A_j(x).$
In the
following we treat the part of the action having to do with
the $xy$ plane, i.e. we consider a two dimensional model; 
generalization to include the remaining hyperplanes is
straightforward. In addition, we consider for both the gauge and the 
scalar field part of the action {\em two} versions of the improvement:
the ``continuum" and the lattice version. In the former
case, a lattice with
spacing equal to $a$ is embedded in the continuum space-time and objects
resembling the usual lattice quantities are considered. An interpolation
is used, which makes it easy to eliminate all of the $a^2$ 
subleading terms both 
in the gauge and the scalar field sectors. The lattice approach is the
treatment of the actual lattice model. It is not in general possible to
eliminate all of the $a^2$ terms and one should be content with a partial
cancellation.

\subsubsection{Lattice embedded in the continuum}
We consider the quantities:
$$
C_{11} \equiv \int_{-\f{1}{2}}^{+\f{1}{2}} dt A_x(x+a t,y-\f{a}{2})
+\int_{-\f{1}{2}}^{+\f{1}{2}} dt A_y(x+\f{a}{2},y+a t)
$$
$$
- \int_{-\f{1}{2}}^{+\f{1}{2}} dt A_x(x+a t,y+\f{a}{2})
- \int_{-\f{1}{2}}^{+\f{1}{2}} dt A_y(x-\f{a}{2},y+a t), 
$$ 
$$ 
C_{12} \equiv \int_{-1}^{+1} dt A_x(x+a t,y-\f{a}{2})
+\int_{-\f{1}{2}}^{+\f{1}{2}} dt A_y(x+a,y+a t)
$$
$$
-\int_{-1}^{+1} dt A_x(x+a t,y+a)
-\int_{-\f{1}{2}}^{+\f{1}{2}} dt A_y(x-a,y+a t),
$$
$$
C_{21} \equiv \int_{-\f{1}{2}}^{+\f{1}{2}} dt A_x(x+a t,y-a)
+\int_{-1}^{+1} dt A_y(x+\f{a}{2},y+a t)
$$
$$
- \int_{-\f{1}{2}}^{+\f{1}{2}} dt A_x(x+a t,y+a)
- \int_{-1}^{+1} dt A_y(x-\f{a}{2},y+a t).
$$ 
Notice that $C_{11}$ represents a continuum version of the $1 \times 1$ 
plaquette, while the two terms $C_{12},~C_{21}$ represent the $1 \times 2$ and 
$2 \times 1$ plaquettes. 

One then expands these quantities in powers of $a$ and ends up with:
\be
a^{-1} C_{11} \simeq [\p_x A_y-\p_y A_x]
+\f{1}{24} a^2 [\p_{xxx} A_y-\p_{yyy}A_x +\p_{xyy} A_y -\p_{xxy} A_x]
\ee
\be
a^{-1} C_{12} \simeq 2 [\p_x A_y-\p_y A_x]
+\f{1}{12} a^2 [4 \p_{xxx} A_y-\p_{yyy}A_x +\p_{xyy} A_y -4 \p_{xxy} A_x]
\ee
\be
a^{-1} C_{21} \simeq 2 [\p_x A_y-\p_y A_x]
+\f{1}{12} a^2 [\p_{xxx} A_y-4 \p_{yyy}A_x +4 \p_{xyy} A_y -\p_{xxy} A_x]
\ee
The terms that will appear in the action can be written more simply 
by using the notations:
$F \equiv \p_x A_y-\p_y A_x,~P \equiv \p_{xxx} A_y-\p_{yyy}A_x 
+\p_{xyy} A_y -\p_{xxy} A_x.$ They read, up to $O(a^2):$
\be
a^{-2} (C_{11})^2 \simeq F^2+\f{1}{12} a^2 F P,~~a^{-2} (C_{12})^2 +
a^{-2} (C_{21})^2 \simeq 8 F^2 +\f{5}{3} a^2 F P.
\ee
Now it is easy to write down the expression for the action up to
$O(a^2):$
$$ 
a^{-2} S = \sum_{x,y} [A a^{-2} (C_{11})^2+B a^{-2} \{(C_{12})^2+(C_{21})^2\}] 
$$
\be
\simeq \sum_{x,y} [A (F^2+\f{1}{12} a^2 F P)+B(8 F^2 +\f{5}{3} a^2 F P)].
\ee
We observe that in general we get the continuum action 
plus the $F P$ terms, which are lattice artifacts. If our aim is to 
better approach the continuum action, we should arrange that 
these artificial terms vanish; in addition, the coefficient of the $F^2$ 
term should be one.
This is easy in this approach: we just choose $A=\frac{5}{3}, ~~ 
B=-\frac{1}{12}.$
Thus the improved action reads:
$$ S=\sum_{x, y, \mu<\nu} [\frac{5}{3} (C_{11}^{\mu \nu}(x,y))^2
-\frac{1}{12} (C_{12}^{\mu \nu}(x,y))^2
-\frac{1}{12} (C_{21}^{\mu \nu}(x,y))^2]$$

\subsubsection{Actual lattice formulation}

In actual lattice calculations one cannot use the interpolation
of the previous section, namely the one based on the t--integrations. One
has link variables and the plaquettes used in the action are sums of four
(or six) such variables. It is therefore very interesting to find out 
how the above analysis is modified if the real situation on the lattice 
is considered.  We will use the same plaquettes as above and Taylor expand 
in powers of $a.$  
To be specific we note that the $1 \times 1$ plaquette is the sum:
\be
 A_x(x,y-\frac{a}{2})+A_y(x+\frac{a}{2},y)
-A_x(x,y+\frac{a}{2})-A_y(x-\frac{a}{2},y)
\ee
while the $2 \times 1$ and $1 \times 2$ plaquettes are the sums:
$$ 
 A_x(x,y-a)+A_y(x+\frac{a}{2},y-\frac{a}{2})+A_y(x+\frac{a}{2},y+\frac{a}{2})
$$ 
\be
-A_x(x,y+a)-A_y(x-\frac{a}{2},y+\frac{a}{2})-A_y(x-\frac{a}{2},y-\frac{a}{2}),
\ee
and
$$ 
 A_x(x-\frac{a}{2},y-\frac{a}{2})+A_x(x+\frac{a}{2},y-\frac{a}{2})
+A_y(x+a,y)
$$ 
\be
-A_x(x-\frac{a}{2},y+\frac{a}{2})-A_x(x+\frac{a}{2},y+\frac{a}{2})
-A_y(x-a,y)
\ee
respectively.
The result of the expansion in $a$ is:
$$
a^{-1} P_{11} = (\partial_x A_y-\partial_y A_x) +\frac{a^2}{24} (\partial_{xxx} A_y
-\partial_{yyy} A_x) + O(a^4)
$$
The results for the two remaining plaquettes are:
$$
a^{-1} P_{12} = 2 (\partial_x A_y-\partial_y A_x) +\frac{a^2}{12} (\partial_{xxx} A_y
+ 3 \partial_{xyy} A_y -4 \partial_{yyy} A_x) + O(a^4)
$$
$$
a^{-1} P_{21} = 2 (\partial_x A_y-\partial_y A_x) +\frac{a^2}{12} (-\partial_{yyy} A_x
- 3 \partial_{xxy} A_x +4 \partial_{xxx} A_y) + O(a^4)
$$
It is straightforward to verify the following:
$$
a^{-2} P_{11}^2 = F^2 + \frac{a^2}{12} (-(\p_{xx} A_y)^2-(\p_{yy} A_x)^2 
+\p_{xy} A_x \p_{xx} A_y+\p_{xy} A_y \p_{yy} A_x) + O(a^4)
$$ 
$$ 
a^{-2} (P_{12}^2+P_{21}^2) = 8 F^2 
-\frac{5 a^2}{3} [(\partial_{xx} A_y)^2+(\partial_{yy} A_x)^2] + O(a^4)
$$ 
$$ 
+\frac{8 a^2}{3} [ (\partial_{xx} A_y)(\partial_{xy} A_x)
+(\partial_{yy} A_x)(\partial_{xy} A_y)]
$$ 
\be
-a^2 [(\partial_{xy} A_x)^2+(\partial_{xy} A_y)^2]
\ee
Now we may consider the linear combination found above and see 
what is the outcome:
$$
a^{-2} S=\frac{5}{3} a^{-2} P_{11}^2 -\f{1}{12} a^{-2} (P_{12}^2+P_{21}^2),
$$
with the Taylor expansion:
\be
F^2 + \f{a^2}{12} [(\p_x F)^2+(\p_y F)^2]
\ee
\be
+\f{29 a^2}{18} [(\p_{xx} A_y)(\p_{xy}A_x)+(\p_{yy} A_x)(\p_{xy} A_y)
-(\p_{xx} A_y)^2-(\p_{yy} A_x)^2] + O(a^4)
\ee

We observe that several $a^2$ terms do not vanish with this 
(or any other) choice of parameters. We have a difficulty, 
stemming from the nature 
of the actual lattice expression of the gauge fields. A possibility to 
eliminate these unwanted terms might be to employ further different 
kinds of plaquettes; however, what really happens is that, 
when bigger plaquettes are considered, new terms appear that cannot 
vanish against existing terms.

An interesting remark is that one may choose different coefficients from 
the ones based on \cite{weisz} and find nicer expressions on the right
hand side:
$$
-\f{1}{11} P_{11}^2+\f{3}{22} (P_{12}^2+P_{21}^2) 
= F^2-\f{3 a^2}{22} [(\p_x F)^2+(\p_y F)^2].
$$

We have decided to use the standard 
non-compact Wilson action (with no improvement) for the gauge part, mainly
because of this difficulty and the consideration of the fact that we intend 
to use the action in the weak gauge coupling regime, so the subleading terms
are not expected to be too serious. 

\subsection{The gauge-scalar part}

We now go ahead with the gauge-scalar sector of the action. The part that 
needs improvement is of course the kinetic term, the only term involving
derivatives. We feel that it is important to improve this part mainly, 
since its big autocorrelation times make mandatory a quicker 
approach to the continuum limit.
Following the scheme  of the previous subsection, we first consider the lattice embedded in a continuous space time and afterwards we turn to the actual 
problem that we face on the lattice.

\subsubsection{Lattice embedded in the continuum}

We start by writing down the continuum kinetic term 
in the ${\hat \mu}$ direction for the scalar field:
\be
\phi^* (\p_\mu-i A_\mu)^2 \phi +h.c. = 
\phi^{* \pp} \phi-\phi^* \phi A^2-i\phi^* \phi A^\pr
-2 i \phi^{* \pr} \phi A +h.c.,
\label{contact}
\ee
where the primes denote differentiations with respect to $x_\mu$ and by $A$ 
we understand $A_\mu.$

The kinetic term in the continuum involves the expression:
\be
\phi^*(x) P e^{\int_x^{x+a {\hat \mu}} A_\mu dx^\mu} 
\phi(x+a {\hat \mu}) + h.c.
\ee
Thus we choose to approximate this kinetic term by the expression:
\be
S_{h1} \equiv \phi^*(x) e^{i a \int_0^1 dt A(x+a t {\hat \mu})} 
\phi(x+a {\hat \mu}) +h.c.
\label{1wwscalar}
\ee
and Taylor expand it in powers of $a.$ The result is found to be:
$$ 
\phi^{* \pp} \phi-\phi^* \phi A^2-i\phi^* \phi A^\pr
-2 i \phi^{* \pr} \phi A
$$ 
$$ 
+a^2(-\f{1}{3} \phi^{*} \phi A A^{\pp}-\f{1}{12} \phi^{*} \phi A^{4}
    -\f{1}{4} \phi^{*} \phi A^{\pr 2} +\f{i}{2} \phi^{*} \phi A^2 A^{\pr})
$$ 
$$ 
+a^2(-\f{i}{3} \phi^{* \pr} \phi A^{\pp}+\f{i}{3} \phi^{* \pr} \phi A^3
    -\f{i}{12}\phi^{*} \phi A^{\ppp} - \f{i}{3} \phi^{* \ppp} \phi A
    -\f{i}{2}\phi^{* \pp} \phi A^\pr)
$$  
$$ 
+a^2(-\f{1}{2} \phi^{* \pp} \phi A^2 - \phi^{* \pr} \phi A A^\pr
    +\f{1}{12} \phi^{* \pppp} \phi)  + O(a^4) +h.c.
$$
As we would like to eliminate the subleading terms, we can add next-to-nearest 
neighbour terms with suitable coefficients. We consider only up to second 
neighbours, that is we consider terms of the form: 
\be
S_{h2} \equiv \phi^*(x) e^{i a \int_0^2 dt A(x+a t {\hat \mu})} 
\phi(x+2 a {\hat \mu}) +h.c.
\label{wwscalar}
\ee
The result of the Taylor expansion contains terms
qualitatively similar to the previous ones:
$$
4 (\phi^{* \pp} \phi-\phi^* \phi A^2-i\phi^* \phi A^\pr
-2 i \phi^{* \pr} \phi A)
$$
$$
+16 a^2(-\f{1}{3} \phi^{*} \phi A A^{\pp}-\f{1}{12} \phi^{*} \phi A^{4}
    -\f{1}{4} \phi^{*} \phi A^{\pr 2} +\f{i}{2} \phi^{*} \phi A^2 A^{\pr})
$$
$$
+16 a^2(-\f{i}{3} \phi^{* \pr} \phi A^{\pp}+\f{i}{3} \phi^{* \pr} \phi A^3
    -\f{i}{12}\phi^{*} \phi A^{\ppp} - \f{i}{3} \phi^{* \ppp} \phi A
    -\f{i}{2}\phi^{* \pp} \phi A^\pr)
$$ 
$$
+16 a^2(-\f{1}{2} \phi^{* \pp} \phi A^2 - \phi^{* \pr} \phi A A^\pr
    +\f{1}{12} \phi^{* \pppp} \phi) + O(a^4) +h.c.
$$
It is easily seen that it is possible to choose the coefficients 
such that the $a^2$ subleading terms vanish. One need only consider
the combination $+\f{4}{3} S_{h1}-\f{1}{12} S_{h2}.$
It is trivial to check that the result for the Taylor expansion 
of this combination reads:
$$
\phi^{* \pp} \phi-\phi^* \phi A^2-i\phi^* \phi A^\pr
-2 i \phi^{* \pr} \phi A + O(a^4) +h.c.,
$$
which is, actually the continuum action (\ref{contact}).

\subsubsection{Actual lattice formulation}

As in the previous case, on the lattice we don't have exactly the
forms (\ref{1wwscalar},\ref{wwscalar}) for the kinetic terms. 
The scalar field kinetic term before
improvement reads: $\sum_{x,{\hat \mu}} \phi^{*}(x) U_{x {\hat \mu}} 
\phi(x+a {\hat \mu}) +h.c.$
To begin with, we write down the Taylor expansion 
of the expression $S_{h1}^{latt} \equiv \phi^{*}(x) e^{i a A_\mu(x+
\f{a}{2} {\hat \mu})} \phi(x+a {\hat \mu}) +h.c.$ The result is:
$$ 
\phi^{* \pp} \phi-i \phi^{*} \phi A^\pr -2 i \phi^{* \pr} \phi A
-\phi^{*} \phi A^2+h.c.
$$
$$
+a^2 (-\f{1}{4} \phi^{*} \phi A^{\pr 2}-\f{1}{12} \phi^{*} \phi A^4
+\f{1}{3} i \phi^{* \pr} \phi A^3)
$$
$$
+a^2 (\f{1}{2} i \phi^{*} \phi A^2 A^{\pr}-\f{1}{2} \phi^{* \pp} \phi A^{2} 
-\phi^{* \pr} \phi A A^{\pr})
$$
$$
+a^2 (-\f{1}{3} i \phi^{* \ppp} \phi A +\f{1}{12} \phi^{* \pppp} \phi
-\f{1}{2} i \phi^{* \pp} \phi A^{\pr})
$$
\be
+a^2 (-\f{1}{4} \phi^{*} \phi A^{\pp} A
-\f{1}{4} i \phi^{* \pr} \phi A^{\pp} -\f{1}{24} i \phi^{*} \phi A^{\ppp} )
 + O(a^4) +h.c.
\label{1neighbor}
\ee
We see immediately that we have terms of order $a^2,$ which we would like to
discard by the improved action. Our step towards the improvement 
will be to consider the next-to-nearest neighbor terms, namely 
$\sum_{x,{\hat \mu}} \phi^{*}(x) U_{x {\hat \mu}}
U_{x+{\hat \mu},{\hat \mu}} \phi(x+2 a {\hat \mu}) +h.c.$

In the following lines we give the Taylor expansion of the expression
$S_{h2}^{latt} \equiv \phi^{*}(x) e^{i a A_\mu(x+\f{a}{2} {\hat \mu})} 
e^{i a A_\mu(x+\f{3 a}{2} {\hat \mu})} \phi(x+ 2 a {\hat \mu}) +h.c. :$
$$
4 \phi^{* \pp} \phi- 4 i \phi^{*} \phi A^\pr -8 i \phi^{* \pr} \phi A
-4 \phi^{*} \phi A^2
$$
$$
+a^2 (-4 \phi^{*} \phi A^{\pr 2}-\f{4}{3} \phi^{*} \phi A^4
+\f{16}{3} i \phi^{* \pr} \phi A^3) 
$$ 
$$
+a^2 (8 i \phi^{*} \phi A^2 A^{\pr}-8 \phi^{* \pp} \phi A^{2} 
-16 \phi^{* \pr} \phi A A^{\pr})
$$
$$
+a^2 (-\f{16}{3} i \phi^{* \ppp} \phi A +\f{4}{3} \phi^{* \pppp} \phi
-8 i \phi^{* \pp} \phi A^{\pr})
$$
\be
+a^2 (-5 \phi^{*} \phi A^{\pp} A
-5 i \phi^{* \pr} \phi A^{\pp} -\f{7}{6} i \phi^{*} \phi A^{\ppp})
 + O(a^4) +h.c. 
\label{2neighbor}
\ee
The next natural step to be taken is, of course, to use the 
coefficients found before and see whether the $a^2$ 
subleading contributions go away or not. 
The result for the linear combination $+\f{4}{3} S_{h1}^{latt} 
-\f{1}{12} S_{h2}^{latt}$ is:
\be
\phi^{* \pp} \phi- i \phi^{*} \phi A^\pr -2 i \phi^{* \pr} \phi A
-\phi^{*} \phi A^2
+a^2 (\f{1}{12} \phi^{*} \phi A^{\pp} A+\f{1}{12} i \phi^{* \pr} \phi A^{\pp}
+\f{1}{24} i \phi^{*} \phi A^{\ppp} )
\ee
We observe that the $a^2$ contributions do not vanish 
completely; the terms in the last lines of equations (\ref{1neighbor}) 
and (\ref{2neighbor}) survive. However, we observe that 
a good deal of the $a^2$ contributions (9 terms out of 12), 
present in the non-improved expression
(\ref{1neighbor}) have disappeared. The conclusion is that we don't really
manage to get rid of all the subleading contributions of order $a^2,$
but we expect that we approach closer to the continuum limit, so,
presumably, the numerical behaviour of the improved action should be better;
this has been confirmed by the simulations.

Thus, gathering everything together, we conclude that the improved action
reads:
$$S=\beta _{g}\sum _{x}\sum _{0<i<j} F_{ij}^{2}
+\beta _{h}\sum _{x}\sum _
{0<i}[\frac{4}{3} (\varphi ^*(x)\varphi (x) 
- \varphi^*(x)U_{i}(x)\varphi (x+\hat i))
$$
$$
- \frac{1}{12} 
(\varphi ^*(x)\varphi (x) - 
\varphi ^*(x)U_{i}(x)U_{i}(x+ \hat i)\varphi (x+2\hat i) )]
$$
\be
+\sum _{x}[(1-2\beta _{R}-3 \f{5}{4} \beta _{h})\varphi ^*(x)\varphi (x)
+\beta_{R}(\varphi ^*(x)\varphi (x))^{2}] 
\label{action}
\ee
where $F_{ij}=\Delta _{i}^f A_{j}(x)-\Delta _{j}^f A_{i}(x)$.

The lattice parameters and the (three--dimensional) continuum ones 
are related as follows:
\be
\beta_{g}=\frac{1}{ag_{3}^{2}}
\ee

\be
\beta_{R}=\frac{x\beta _{h}^{2}}{4\beta _{g}} 
\label{betar}
\ee
\be
2\beta_g^{2} \frac{1-3 \frac{5}{4} \beta _{h} -2\beta_R}{\beta_h}=
y-(1+4x)\frac{\Sigma^\prime\beta _{g}}{4\pi }
-\frac{\Sigma \beta_{g}}{4\pi}-\frac{\beta_{g}}{12}
\label{basic}
\ee
where $\Sigma=3.176$ and $\Sigma^\prime=2.752.$
In the appendix we prove the relation (\ref{basic}).

\section{Results}
We used the Metropolis algorithm for the updating of both the gauge and
the Higgs field. It is known that the scalar fields have much longer
autocorrelation times than the gauge fields. Thus, special care must be
taken to increase the efficiency of the updating for the Higgs field. We
made the following additions to the Metropolis updating procedure 
\cite{klrs}:

{\bf a) Global radial update:} We update the radial part of the Higgs field
by multiplying it by the same factor at all sites: $R(\vec x)
\rightarrow e^{\xi }R(\vec x),$ where $\xi \in [-\varepsilon ,
\varepsilon]$ is randomly chosen. The quantity $\varepsilon$ is adjusted
such that the acceptance rate is kept between 0.6 and 0.7. The probability
for the updating is $P(\xi )=$ min$\{1,\exp (2V\xi -\Delta S(\xi)) \}$ where
$\Delta S(\xi )$ is the change in action, while the $2V\xi$ term comes
from the change in the measure.

{\bf b) Higgs field overrelaxation:} We write the Higgs potential at
$\vec x$ in the form: 
\be
V(\varphi (\vec x))=-{\bf a} \cdot {\bf F} +
R^{2}(\vec x)+\beta _{R}(R^{2}(\vec x)-1)^{2} 
\ee
where
$${\bf a} \equiv \left( \begin{array}{c} R(\vec x) \cos \chi (\vec x)\\
                                   R(\vec x) \sin \chi (\vec x)
                   \end{array}
                               \right),$$
$$
{\bf F} \equiv \left(\begin{array}{c} F_1\\F_2 \end{array}
\right),$$
$$ F_1 \equiv \beta_h \sum_i [\frac{4}{3}R(\vec x+ \hat i)\cos
(\chi (\vec x+\hat i)+\theta(\vec x))-\frac{1}{12} 
R(\vec x+2 \hat i)\cos
(\chi (\vec x+2\hat i)+\theta(\vec x)+\theta(\vec x +\hat i))], 
$$
$$
F_2 \equiv \beta_h \sum_i [\frac{4}{3} 
R(\vec x+\hat i)\sin (\chi (\vec x+\hat i)+\theta(\vec x))
-\frac{1}{12} R(\vec x+2\hat i)\sin (\chi (\vec x+2\hat i)
+\theta(\vec x)+\theta(\vec x +\hat i))].
$$
We can perform the change of variables: $({\bf a},{\bf F}) \rightarrow
(X,F,{\bf Y})$ ,where
\be
F \equiv |{\bf F}|,~~~ {\bf f} \equiv \frac{{\bf F}}{\sqrt{F_1^2 + F_2^2}},~~~
X \equiv {\bf a} \cdot {\bf f},~~~{\bf Y} \equiv {\bf a} - X {\bf f}.
\ee

The potential may be rewritten in terms of the new variables:
\be
\bar V(X,F, {\bf Y})=-XF +(1+2\beta _{R}({\bf Y}^{2}-1)) X^{2}
+{\bf Y}^{2}(1-2\beta_{R})+\beta _{R}(X^{4}+{\bf Y}^{4}).
\ee
The updating of ${\bf Y}$ is done simply by the reflection:
\be
{\bf Y} \rightarrow {\bf Y}'= -{\bf Y}.
\ee

The updating of X is performed by solving the equation:
\be
\bar V(X^\pr,F, {\bf Y})= \bar V(X,F, {\bf Y})
\ee
with respect to $X^\pr.$
Noting that $X^\pr=X$ is obviously a solution, we may factor out the 
quantity $X^\pr - X$ and reduce the quartic equation into a cubic one, which
may be solved.
The change $X \rightarrow X'$ is accepted with probability:
 $P(X')=$ min$\{P_0,1\},$ where $P_0 \equiv
 \frac{\partial \bar V(X,F,{\bf Y})}{\partial X}/\frac{\partial \bar
 V(X',F',{\bf Y'})}{\partial X'}$.

For our Monte--Carlo simulations we used cubic lattices with 
volumes $V=8^3,12^3,16^3$. 
For each volume we performed 30000--50000 
thermalization sweeps and 60000--100000 measurements.
We have set the value of $x$ equal to 0.0463. According to the relation (10)
using $m_{W}=80.6 GeV$ and $g=\frac{1}{3}$ this value of $x$ corresponds
to Higgs field mass $m_{H}=30 GeV$. For each value of $\beta_{h}$ we determine 
the value of $\beta_{R}$ by using relation (\ref{betar}).
The phase transition is expected to be of first order for such a low mass of
the scalar field.

We have used five quantities to determine the phase transition points:

\begin{enumerate}

\item The distribution $N(E_{link})$ of $E_{link}.$

\item The susceptibility of $E_{link} \equiv \frac{1}{3V}
\sum_{x,i} \Omega^*(x)U_{i}(x) \Omega(x+i)$ (we have set $\varphi(x) \equiv
R(x) e^{i \chi(x)} \equiv R(x) \Omega(x)$ ):
$$S(E_{link}) \equiv V (<(E_{link})^{2}>-<E_{link}>^{2}).$$

\item The susceptibility of $R2 \equiv \frac{1}{V} \sum_x R^2(x):$
$$S(R2) \equiv V (<(R2)^2>-<R2>^2).$$

\item The Binder cumulant of $E_{link}$:
$$ C(E_{link})=1-\frac{<(E_{link})^4>}{3 <(E_{link})^2>^2}.$$

\item The Binder cumulant of $R \equiv \frac{1}{V} \sum _x R(x)$:
$$ C(R)=1-\frac{<(R)^4>}{3 <(R)^2>^2}.$$  

\end{enumerate}

The pseudocritical $\beta_h^*(A,V)$ values have been found by determining 
(a) equal heights of the
two peaks of the distribution $N(E_{link}),$ (b) the maxima of the
quantities $S(E_{link}),~S(R2)$ and (c) the minima of the cumulants
$C(E_{link}),C(R).$ 
The values $\beta_h^*(A,V)$ depend on the specific quantity (denoted by A)
which has been employed, as well as on the volume V. It has to be noticed that
while searching we have made use of the Ferrenberg--Swendsen reweighting
technique (\cite {fs}) to find the pseudocritical $\beta_{h}$ for the volume
$16^3$. 

In figure 1 we depict the behaviour of the susceptibility $S(E _{link})$ versus
$\beta _{h}$ for three lattice volumes. The curves are fitted through the data;
for simplicity we give the actual measurements for the largest volume. The curves
represent the data quite nicely. In calculating the error bars we first found
the integrated autocorrelation times $\tau _{int}(A)$ for the relevant quantities
A and constructed samples of data separated by a number of steps greater than 
$\tau _{int}(A)$. Then the errors have been calculated by the Jacknife method 
(\cite {jack}), using the samples constructed according to the procedure just
described. Notice that the peak values increase almost linearly with the
volume which is characteristic of a first order transition. 

\begin{figure}
\centerline{\hbox{\psfig{figure=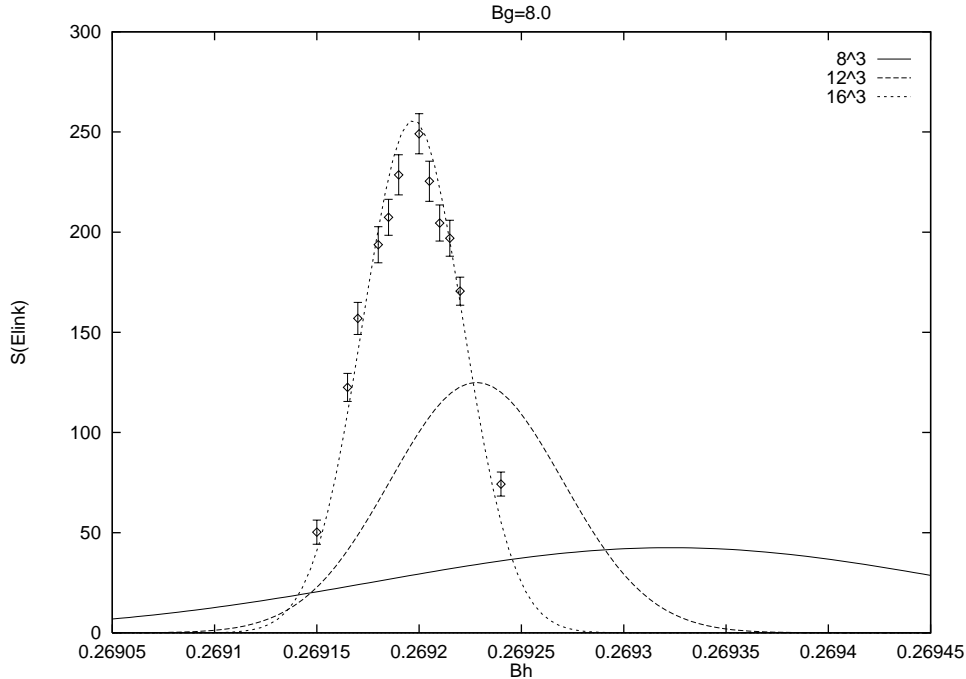,height=9cm,angle=-90}}}
\caption[f1]{Susceptibility of $E_{link}$}
\label{f1}
\end{figure}

In figures 2 and 3 we depict the behaviour of the Binder cumulants $C(E_ {link})$ and
$C(R)$ for three lattice sizes. Again, we show the real measurements for the
largest volume. The error bars have been calculated, also, by the Jacknife method.
The volume dependence of the cumulants display evidence for a first order
phase transition.


\begin{figure}
\centerline{\hbox{\psfig{figure=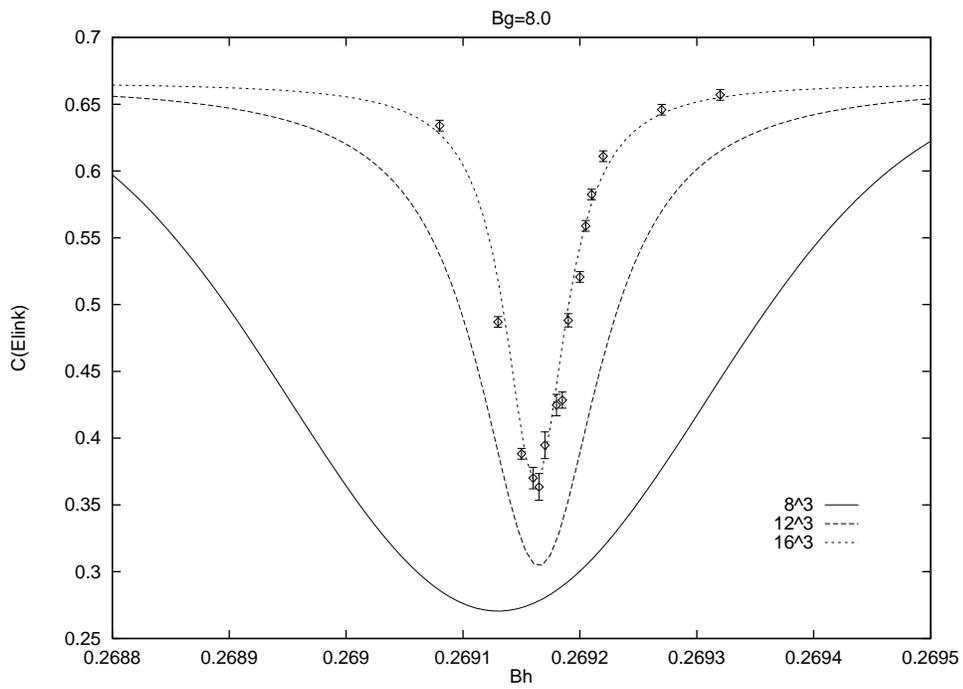,height=9cm,angle=-90}}}
\caption[f2]{Cumulant of $E_{link}$}
\label{f2}
\end{figure}
\begin{figure}
\centerline{\hbox{\psfig{figure=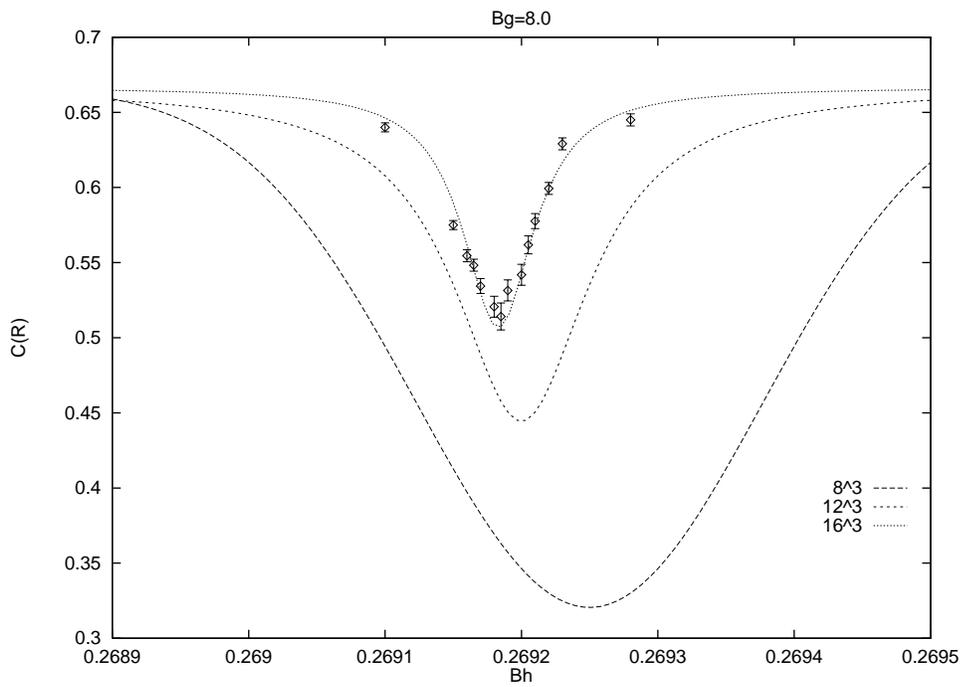,height=9cm,angle=-90}}}
\caption[f3]{Cumulant of $R$}
\label{f3}
\end{figure}

The different values of $\beta_h^*(A,V)$ corresponding to the quantities A
are due to the finite lattices used. So, we should extrapolate these values to 
infinite volume adopting the ansatz:
$$\beta_h^*(A,~V) = \beta_h^{cr}(\infty) +\frac{c(A)}{V},$$

The extrapolated value $\beta_h^{cr}(\infty)$ should not depend on quantity A 
because this is the infinite volume extrapolation for the critical point.

Figure 4  shows the extrapolation  to infinite volume using data for the 
pseudocritical $\beta_h^*(A,V)$ values obtained from the various quantities A
versus the inverse lattice volume, along with the linear fits to the data. The 
error bars in $\beta_h^*(A,V)$ have been calculated from the statistical error 
of the values of the quantities A at the critical point.
One can observe that, at, finite volumes, the smallest pseudocritical values
are given by the cumulant of $E _{link}$; then, in ascending order, the 
values given by the cumulant of $R$, the equal height, the susceptibility of
$E _{link}$ and the susceptibility of $R2$. Also, we notice that the infinite volume 
extrapolation is almost independent from the specific quantity used; the 
differences at the point $\frac{1}{V}=0$ between the various extrapolated values
are less than $10^{-5}$. 
The critical values lie in the interval (0.269176, 0.269183). In our previous publication \cite{dimo} 
where we worked with the same model, but
without any improvement to the Higgs part of the action,
 we had found that for the same value of $\beta_{g}$ the critical
$\beta_{h}$ values were lying in the interval (0.336932, 0.336940). 
Although the precision is comparable, one should notice that the 
result presented here has been found by using three times smaller lattice 
volumes than the previous one which means much shorter
computer time. Evidently, this result is due to the effect of improving 
the lattice Higgs action which provides a quicker approach to 
the thermodynamic limit.

\begin{figure}
\centerline{\hbox{\psfig{figure=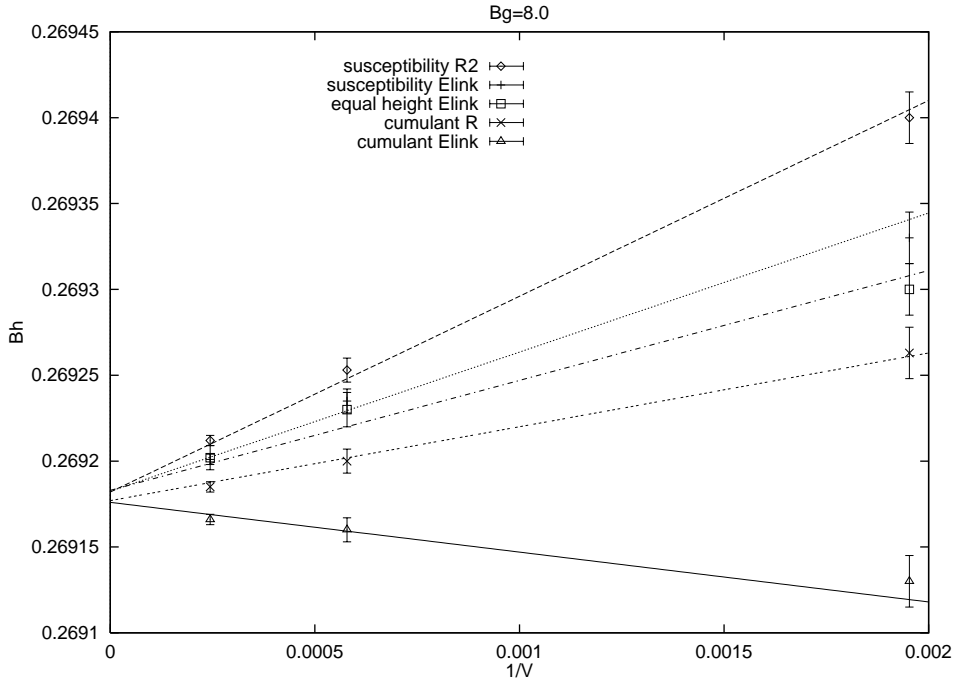,height=9cm,angle=-90}}}
\caption[f4]{Extrapolation for $\beta_g=8$}
\label{f4}
\end{figure}

We can, now, predict the critical temperature $T _{cr}$. Actually, the 
quantity $\beta_h^{cr}$ yields $y _{cr}$ through 
equations (\ref{betar}, \ref{basic}); 
then equation (\ref{yT}) gives $T _{cr}=130.64(19)$. 
In \cite {dimo} using 2--loop calculations it was found
$T _{cr}=131.18(14)$ while a 1--loop calculation would give 
$T _{cr}=130.74(14)$.
Thus, the two 1--loop results are almost identical, but the first one
has been achieved in a more economical way due to the improved action used.

\section*{Acknowledgements}

We would like to acknowledge financial support from the
TMR project ``Finite temperature phase transitions in particle Physics",
EU contract number: FMRX-CT97-0122. 

\newpage

\section{Appendix}

We want to prove the relation (\ref{basic}), 
which relates the masses on the lattice
and in the continuum up to 1--loop perturbation theory. In general we have:
\be
m_{L}^{2}=m_{3}^{2}-\delta m_{L}^{2}(h)
\label{count}
\ee
The counterterm will be calculated by considering the $\frac{1}{a}$ 
terms from the 1--loop lattice effective potential (\cite {laine}).

We consider the pure gauge and gauge fixing part of the action 
(\ref{action}). We use 
the relation $\beta_g=\frac{1}{g_3^2 a}$ and write down the kinetic term as:
$$
S_g = a^3 \sum_x \frac{1}{4} \sum_{i \ne j} 
\left( \frac{\Delta_i^f A_j -\Delta_j^f A_i}{a} \right)^2
$$
Let us comment here a little bit about the gauge fixing term, which is
the new element here. For the perturbative treatment of Higgs models
it is usual to employ the $R_\xi$ gauges. This choice is dictated by its
simplicity, since then the mixing terms between the gauge field and the
would-be Goldstone bosons vanish. However, one should be careful with
this gauge, since the so-called Nielsen identities \cite{nielsen}
should be satisfied \cite{aitch}.
This severely restricts the possible $R_\xi$ gauges; violation of these
identities will lead to unphysical results. This is the reason that
we have chosen to stick to the Lorentz gauge fixing for the gauge action:
$$L_{gf}=\frac{a}{2\xi}\sum_{x}\sum_{i}[A_{i}(x)-A_{i}(x-\hat{i})]^2.$$
This gauge does not face the previous complications, so we prefer to stay
on the safe side, at the price, of course, of having also to deal with the
non-vanishing mixed terms. Note that we take the
Landau gauge $\xi=0$ at the end of the calculations.

Going to the Fourier space (notice that 
$A_{\mu}(x)=\int \f{d^3 p}{(2 \pi)^3} e^{i p(x+\frac{a\hat{\mu}}{2})} 
{\tilde A}_{\mu}(p)~),$ 
we find:
\be
S_g+S_{gf} = 
\frac{1}{2}\int \f{d^3 p}{(2 \pi)^3} 
\sum_{i,j>0}{\tilde A}_{i}(p)[(\tilde{p}^2 
+\frac{1}{a^{2}}(\frac{1}{\xi}-1) \tilde{p}_i \tilde{p}_j] {\tilde A}_j(-p)
\label{act1}
\ee
where
\be
\tilde{p}_{i}=\frac{2}{a} sin \frac{p_{i}a}{2}
\ee

Next we consider the part of the action involving the scalar fields. We
follow essentially the same procedure as above, but a step that should be
taken is the decomposition 
$\phi=\frac{1}{\sqrt{2}}(\phi_{0}+\phi_{1}+i\phi_{2})$ of the scalar field.
Then the rescaling $(\frac{\beta_{h}}{2a})^{\frac{1}{2}}\varphi=\phi$ has
to be performed to get the corresponding part of the action in continuum form.
The Fourier transform of the relevant part of the action reads:
$$ 
S_H = +\frac{1}{2}\int \f{d^3 p}{(2 \pi)^3} 
\phi_{1}(p)(\tilde{p}^2 +\frac{a^2}{12}\tilde{p}^{4}
+m_{1}^{2})\phi_{1}(-p)
$$
$$
+\frac{1}{2}\int \f{d^3 p}{(2 \pi)^3} 
\phi_{2}(p)(\tilde{p}^2 +\frac{a^2}{12}\tilde{p}^{4}
+m_{2}^{2})\phi_{2}(-p)
$$
\be
+\int \f{d^3 p}{(2 \pi)^3} 
{\tilde A}_i(p) (\frac{m_{T}}{\alpha} \tilde g_{i}) \phi_{2}(-p)
+\frac{1}{2}\int \f{d^3 p}{(2 \pi)^3} {\tilde A}_{i}(p)(m_{T}^{2}
+\frac{\alpha^{2}m_{T}^{2}}{12}\tilde p_{i}^{2}) {\tilde A}_{i}(-p)
\label{act2}
\ee
with $\tilde{g}_{i} \equiv \frac{5}{4}\tilde{p}_{i}-\frac{1}{12}\hat{p}_{i}.$
Let us collect some additional notations that have just been used:
\be
\hat{p}_{i} \equiv \frac{2}{a} sin \frac{3 p_{i}a}{2},~
\tilde{p}^{2} \equiv \sum_{i}\tilde{p}_{i}^{2},~
\tilde{p}^{4} \equiv \sum_{i}\tilde{p}_{i}^{4},~
\ee
\be
m_{T}^2 \equiv g_{3}^2 \phi_{0}^2,~ m_{1}^{2} \equiv 
m_{3}^{2}(\mu)+3\lambda_{3}\phi_{0}^{2},~
m_{2}^{2} \equiv m_{3}^{2}(\mu)+\lambda_{3}\phi_{0}^{2}.
\label{masses}
\ee
Now, considering the quadratic part of the action, which is contained in the
above equations, it is easy to read out the propagators $D_1,D_2$ of the fields 
$\phi_{1},\phi_{2}$ respectively:
\be
D_{1}^{-1}=\tilde{p}^2 +\frac{a^2}{12}\tilde{p}^{4}+m_{1}^{2},~~
D_{2}^{-1}=\tilde{p}^2 +\frac{a^2}{12}\tilde{p}^{4}+m_{2}^{2}.
\ee
In the equation above the terms proportional to $\tilde{p}^{4}$ arise 
directly from the improvement terms in the scalar sector 
concerning the next-to-nearest-neighbour contribution.

The effective potential at one loop is found using the relation:
$$
V_{L}^{1-loop}=-ln(Z),
$$
where $Z \equiv \int [D \phi_1][D\phi_2][DA]  e^{-S}.$ Keeping only 
the quadratic part of the action we have just Gaussian integrations, so 
we easily get the result:
$$
V_{L}^{1-loop} = \frac{1}{2}\int \f{d^3 p}{(2 \pi)^3} ln(D_{1}^{-1}) 
+\frac{1}{2}\int \f{d^3 p}{(2 \pi)^3} ln(D_{2}^{-1})+
$$
\be
\frac{1}{2}\int \f{d^3 p}{(2 \pi)^3} ln(det(\Delta_{ij}^{-1}-N_{ij}))
\label{3term}
\ee
where
\be
  \Delta_{ij}^{-1}-N_{ij}=(\tilde{p}^{2}+m_{T}^{2}+\frac{a^2 m_{T}^{2}}{12}
  \tilde{p}_{i} \tilde{p}_{j}) \delta _{ij} +\frac{1}{a^2}(\xi^{-1}-1)
  \tilde{p}_{i} \tilde{p}_{j}- \frac{m_{T}^{2}}{a^2} \frac{ \tilde{g}_{i}
   \tilde{g}_{j}}{D_{2}^{-1}},
\ee

We write the gauge propagator in this form to display the
contribution $N_{ij} \equiv \frac{m_{T}^{2}}{a^2} \frac{ \tilde{g}_{i}
   \tilde{g}_{j}}{D_{2}^{-1}}$, which is due to the mixing term between 
$A_{i}$ and the imaginary part of the Higgs field.

An integral which appears very frequently and should be calculated is 
the following:
\be
I(m)=\frac{1}{2}\int_{-\pi/a}^{\pi/a} \frac{d^{3}p}{(2\pi)^{3}}ln(\tilde{p}^2 +
\frac{a^2}{12}\tilde{p}^{4}+m^{2})
\ee
 
If we differentiate the above with respect of $m$ we take:

\be
dI(m)=mK(m)dm
\label{diffe}
\ee

where 
\be
aK(m)=\int_{-\pi}^{\pi}\frac{d^{3}p}{(2\pi)^{3}} \frac{1}{\bar{p}^{2}
+\frac{1}{12}\bar{p}^{4}+M^{2}}
\label{Kintegral}
\ee
with:
$M^{2}=(am)^{2},$ $\bar{p}^{2}=4\sum_{i}sin^{2}\frac{p_{i}}{2}$ and $\bar{p}^{4}=16\sum_{i}sin^{4}
\frac{p_{i}}{2}$
At this point we follow (\cite {Moore}) where the expansion of 
(\ref{Kintegral}) in powers of $M=am$ is to be used.
In the following we denote  by $B$ the 
``Brillouin zone" $[-\pi,+\pi]^3,$ 
 in addition, 
$\bar{\Pi}^2 \equiv \bar{p}^{2}+\frac{1}{12}\bar{p}^{4}.$ 
Then the following equality holds:
$$
\int_B \f{d^3 p}{(2 \pi)^3} \f{1}{\bar{p}^{2}+\frac{1}{12}\bar{p}^{4}+M^2 } = 
\int_B \f{d^3 p}{(2 \pi)^3} \f{1}{(\bar{\Pi}^2)^2}
-\int_B \f{d^3 p}{(2 \pi)^3} \f{M^2}{\bar{\Pi}^2 (\bar{\Pi}^2 +M^2)}
$$
The first integral equals $\f{\Sigma^\pr}{4 \pi};$ the 
contribution of the second integral can be shown to be of $O(a),$ so it 
will not contribute to the infinite part.
Gathering everything together yields:
$$
K(m)=\frac{\Sigma^{\prime}}{4\pi a}+
{\rm finite}
$$
Hence, integrating (\ref{diffe}) we get: 
\be
I(m)=\frac{1}{2}\frac{\Sigma^{\prime}}{4\pi a} m^{2}+
{\rm finite}
\label{effpot}
\ee
where 
\be
\Sigma^{\prime}=\frac{1}{(\pi)^{2}}\int_{0}^{\pi}d^{3}p 
\frac{1}{\sum_i (sin^2 \f{p_i}{2} +\f{1}{3} sin^4 \f{p_i}{2}})
\ee
is being calculated numerically at the value of $\Sigma^{\prime}=2.752.$ 

Up to this point we are ready to calculate the infinite part in the 
effective potential (behaving like $\f{1}{a})$ that is due to the scalar 
fields only. The mass terms coming from the fields $\phi_1,~\phi_2$ are:
$$
\f{1}{2} \f{\Sigma^{\prime}}{4\pi a} m_1^2,
~~\f{1}{2} \f{\Sigma^{\prime}}{4\pi a} m_2^2
$$
respectively. Recalling (\ref{masses}), their derivatives with respect to the 
classical field $\phi_0^2$ are:
\be
(3 \lambda_3 + \lambda_3) \f{1}{2} \f{\Sigma^{\prime}}{4\pi a}  = 4 x g_3^2 
\f{1}{2} \f{\Sigma^{\prime}}{4\pi a}
\ee

The next step is to calculate the gauge contribution to the effective 
potential, that is the third term in equation (\ref{3term}). The calculations
are straightforward, but quite tedious; we just mention that we directly 
expand the determinant and keep contributions only up to order $a^2,$ since
the $a^4$ terms will yield merely finite results, which are not our
concern here. We only write down the final result:
$$ 
V_L^{1-loop} = \int \f{d^3 p}{(2 \pi)^3} 
[ln({\tilde p}^2 +m_T^2)^2 +(ln {\tilde p}^2 -ln \xi)
$$
\be
+ln(1+ 2 \f{\f{a^2 m_T^2}{12}}{{\tilde p}^2 +m_T^2} 
\f{{\tilde p}^2_1{\tilde p}^2_2+{\tilde p}^2_2{\tilde p}^2_3
+{\tilde p}^2_3{\tilde p}^2_1}{{\tilde p}^2} ) +O(a^4) ]
\ee
We note that the second term is exactly the same as the one appearing 
in the continuum counterpart of the model, so they cancel upon
comparison with the continuum model. These terms do 
not depend on the mass, so they present no interest for the 
calculation of the effective potential anyway. We should note that 
we have not given the full result of the third term; 
to keep things simple, we gave its value for $\xi = 0.$
The above expression can be written as:
\be
\frac{1}{2}\int_{-\pi/a}^{\pi/a} \frac{d^{3}p}{(2\pi)^{3}}ln(\tilde{p}^2 +
\frac{a^2}{12}\tilde{p}^{4}+m_{T}^{2})
+\frac{1}{2}\int_{-\pi/a}^{\pi/a} \frac{d^{3}p}{(2\pi)^{3}}
ln(\tilde{p}^2 + m_{T}^{2})+\f{1}{2} \f{1}{12a} m_{T}^{2}
\label{mt2}
\ee
The integrals appearing in equation (\ref{mt2}) have already been computed; the
infinite part of the effective potential reads:
\be
\f{1}{2} (\f{\Sigma^\pr}{4\pi a} m_{T}^{2}+\f{\Sigma}{4\pi a} m_{T}^{2}
+\f{1}{12 a} m_{T}^{2})
\ee
We invoke equation (\ref{masses}) and calculate again the second derivative 
with respect to the classical field $\phi_0$ to find out that the 
infinite part reads:
\be
\f{g_3^2}{2} (\f{\Sigma^\pr}{4\pi a}+\f{\Sigma}{4\pi a}+\f{1}{12 a})
\ee
Collecting all the infinite contributions we can match the counterterm
equation and this yields:
$$
2\beta _{g}^{2}(\frac{1}{\beta_{h}}-3 \frac{5}{4}-\frac{2\beta _R}{\beta_h})
= y-(1+4x)\frac{\Sigma^\prime\beta _{g}}{4\pi }
-\frac{\Sigma \beta_{g}}{4\pi}-\frac{\beta_{g}}{12},
$$ 
which is the equation (\ref{basic}).

\end{document}